\documentclass[10pt,a4paper]{article}
\usepackage{fontenc}
\usepackage{calc}
\usepackage{indentfirst}
\usepackage{fancyhdr}
\usepackage{graphicx}
\usepackage{lastpage}
\usepackage{ifthen}
\usepackage{lineno}
\usepackage{float}
\usepackage{amsmath}
\usepackage{setspace}
\usepackage{enumitem}
\usepackage{mathpazo}
\usepackage{booktabs}
\usepackage{titlesec}
\usepackage{etoolbox}
\usepackage{amsthm}
\usepackage{hyphenat}
\usepackage{hyperref}
\usepackage{footmisc}
\usepackage{geometry}
\usepackage{caption}
\usepackage{url}
\usepackage{mdframed}
\usepackage{subcaption}

\usepackage{times}

\title{Making Sense of the World: Framing Models for Trustworthy Sensor-Driven Systems}
\author{Muffy  Calder $^{1}$*, Simon Dobson $^{2}$, 
  Michael  Fisher  $^{3}$, and Julie McCann $^{4}$\\[1em]
  \begin{tabular}{ll}
$^{1}$ & School of Computing Science, University of Glasgow,
    UK\\
    & [corresponding author: \texttt{muffy.calder@glasgow.ac.uk} \\
    $^{2}$ & School of Computer Science, University of St. Andrews, UK \\
    $^{3}$ & Dept of Computer Science,  University of Liverpool,  UK  \\
        $^{4}$ & Dept of Computing, Imperial College London,  UK
      \end{tabular}}
\date{[Accepted for Computers 7(4):62; \texttt{https://doi.org/10.3390/computers7040062 }]}
\begin{document}
 \maketitle

\abstract{Sensor-driven systems provide data and information that facilitate real-time decision-making and autonomous actuation, as well as  enable informed policy choices. But can we be sure these systems work as expected,  can we   model  them  in a way that captures
all the key issues?  We define two concepts: \emph{frames of reference} and \emph{frames of function} that help us organise models of sensor-based systems and their purpose.  Examples from  a smart water distribution network  illustrate  how  frames  offer a lens through which to organise and balance multiple views  of the system. Frames aid communication between modellers, analysts and stakeholders,   and distinguish the purpose  of each model,   which contributes towards our   trust that the system fulfils its purpose.}

 
\section{Introduction}

Sensor-driven systems are everywhere: from transportation and
buildings, to smart tags, power systems and environmental
monitoring. They provide data and information that facilitate
real-time decision-making and autonomous actuation, as well as
enabling informed policy choices by service providers and regulators. 
 The  Internet of Things depends on robust, sensor-driven systems
that can be trusted to deliver useful, correct, and timely
information.  Our vision is of smarter sensor-based systems of which
stakeholders -- from developers to  scientists and policy makers -- can ask deeper
questions while being confident of obtaining reliable answers. But can we be convinced that these systems do what we expect, can their stakeholders ask deep questions and be
confident of obtaining reliable answers? Are they trustworthy with respect to their requirements? This is especially difficult to determine, for example when systems need to balance competing demands such as   sampling  rates required to answer a scientific question, timely actuation to ensure delivery of a service, and energy conservation when operating with constrained resources.

Given the ubiquity of sensor-driven networked systems, it is perhaps surprising
that extracting reliable information from them remains far from
straightforward: sensors consume power, they are noisy, they decalibrate, may become
misplaced, moved, compromised, and generally degraded over time, both
individually and as a collective. This is beyond traditional software
engineering: uncertainty,  failures,  power   and  communication constraints    
pervade  both the physical and digital
environments in which these systems operate, and the sensors themselves, 
and is even crucial to  some of the adaptive algorithms employed. Furthermore, sensors
themselves are increasingly required to fulfil
both housekeeping functions (reporting to the sensor provider) and
multiple sensing functions (reporting to multiple applications). The
sensor system may also be required to be {\em smart}, e.g.~support, and exhibit  increasing
degrees of autonomy, self-awareness, and intelligence.   These systems are core to many technologies intended to aid sustainability via smart cities, smart grids, smart farming etc.~and yet their own reliability and sustainability is not well understood. 
Can we guarantee that the system will do what we expect,
and therefore can we, in good conscience, make use of the information
being presented, and so trust the behaviour being exhibited? Without
such assurances, no business or organisation will deploy complex,
semi-autonomous, adaptive sensor-driven systems; no decision-maker
will allow their decisions to rest on a possibly unstable foundation.

 While we cannot engineer away  issues such as uncertainty, failures, energy consumption etc.~fortunately,  
sensor-driven system architecture   tends to be  specialised and  constrained, usually, to an enhanced form of a  sense and control loop. 
Thus while assurance is still difficult, it is within the context of  specialised  concerns and architecture.   It 
is our assertion that a crucial tool for addressing trustworthiness is
the use of \emph{models}, both at design time and run-time (e.g.~online models): for
specification; for explanation; and for exploration and
prediction.   
Our  assertion derives from our experience  of research and development of a wide range of  sensor-driven
systems e.g.~\cite{kartakis2015waterbox,DobsonACM2014,ucami-13,
FisherDW13,ScatterboxVerification,MAM2015,Savannah,Abnormal16,Jackson:2017}.

\begin{figure}
  \begin{center}
    \includegraphics[width=\columnwidth]{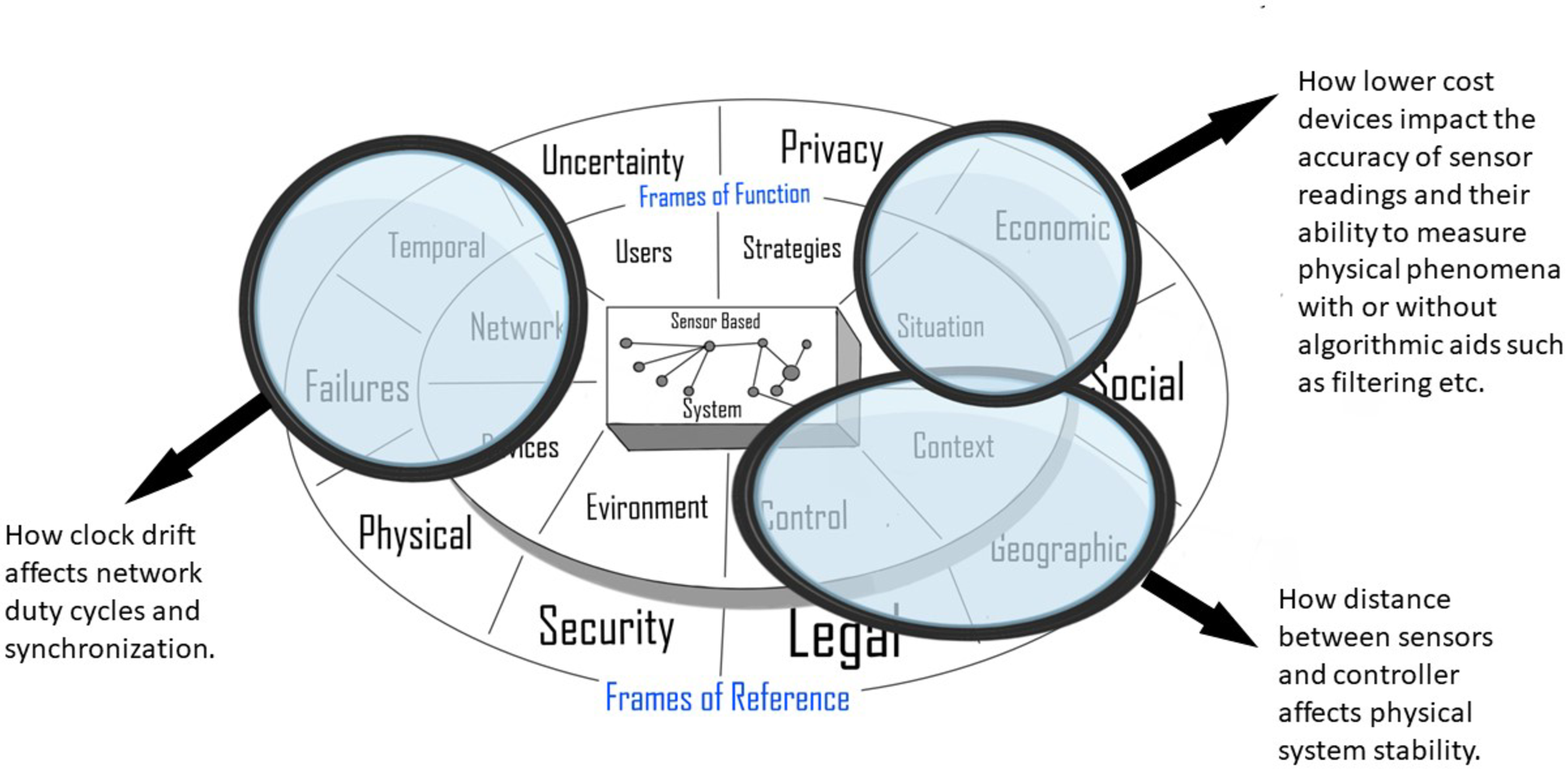}
  \end{center}
  \caption{Frames of reference and function: a lens for modelling and analysis.}
  \label{S4diagram}
\end{figure}

For any sensor-driven system,   many models   \emph{could} be produced;   we define two categorisations that are orthogonal yet overlapping, which 
  help guide us. These two new concepts contribute to a principled modelling process by helping us to frame models of sensor-based systems and their purpose: {\em frames of reference} provide a way to organise and balance multiple perspectives and concerns, and {\em frames of function}     identify  
 component(s) within the system architecture that  are the focus of attention. Together, they offer a lens through which to view  and analyse the sensor-based system, which aids   understanding of the purpose of that model and    communication between modellers, analysts, and stakeholders. This allows us to distinguish and articulate the purpose  of each model,  contributing to our overall
trust that the system fulfils its purpose and requirements. Fig.~\ref{S4diagram} illustrates how different combinations of the  frames can allow the multiple dimensions of focus to be reduced to make analysis more tractable, in terms  the  development of models based on the questions we wish to ask of the system.  
 
The following sections describe in more detail   the two concepts that help us to frame model purpose, questions and analysis.
 In the next section we review  the motivation for  models 
and in  Section~\ref{reference} and Section~\ref{models} we define frames of reference and frames of  function.   In ~\ref{when} we discuss how and when to use  frames in modelling and in 
Section~\ref{example} we illustrate the new concepts by way of     models for a smart   water distribution network.   We offer  a short discussion in Section~\ref{discussion}   and related work is discussed in Section~\ref{related};    conclusions are provided in Section~\ref{conc}. 
It is important to note that this view provides an abstract
methodology for design and development. While this must be subsequently populated with concrete languages and ontologies, these are typically very different for each scenario and so we will not expand on these here.

\section{Sensor-based  systems and their models} 
The concerns of  uncertainty,
failures,  energy etc.~that are inherent in practical sensor-based systems
will \emph{always} be present: these aspects cannot simply be
engineered away and subsequently ignored.  Further,  the purpose, or
mission, of the system may involve a number of trade-offs and
compromises.  A simple example involves controlling the duty cycle of a
sensor node to preserve battery life. Efficient battery technology coupled with low-powered micro-controllers and transceiver devices have been key to the uptake of wireless sensors and now such systems can be placed in more remote and difficult to reach areas with minimal infrastructure required. Yet batteries, from a sensor systems point of view, have become an ``Achilles heel'' for the long term sustainability and operation of wireless sensor systems. It is costly to replace batteries (and problematic if the sensors are inaccessible), they can be dangerous (lithium can cause fires),  contribute to waste, and generally have a negative environmental effect. Further, energy harvesting is not always a solution either since energy storage is inherently inefficient and again storage cells last only for a short number of years (think of mobile phone battery lifetimes). So,  core to making sensor systems `green' is energy management that typically implies at least duty-cycling and possibly adaptive duty-cycling. However, the latter might be constrained by a
mission goal prescribing minimal sensing frequency -- or might even be
disallowed entirely because of the need for extraction of a regular time
series. Without  precisely-articulated requirements, it is impossible to
know how this trade-off can be made in a way that does not compromise
the efficacy of the sensor-based system. It is difficult to answer questions about energy source choice if we have not defined our goals and we do not have a way to explore  design alternatives and their impact. Often such choices are implicit; formal models
make them explicit, and thus amenable to analysis.
Models are  at the core of enabling stakeholders to have the
confidence they need to trust in -- and make decisions from -- the
information being returned from a sensor system.

Models are used extensively across   science and engineering, providing insight
into the behaviours of \emph{what already exists}, predicting future
behaviours of \emph{what might exist}, and exploring the implications
of the specifications and designs of \emph{what we want to
  exist}  and how and if they meet system requirements \cite{Modelling}. We do not repeat the arguments in support of modelling here, but note that 
recent research in pervasive and cyber-physical systems that tackles modelling, implementation, and
 deployment of artificial systems integrated within real-world
 environments,  e.g.~\cite{Tsigkanos,DerlerLV12},  is highly relevant.
We draw on  this work, for
example from pervasive systems~\cite{ContextIsCentral}, where raw data from devices is
processed as \emph{context}, from which \emph{situations} are
inferred, and control and adaptation then depends on those situations, and
 we draw on cyber-physical systems to the extent to which they
concentrate on the modelling and interaction of software with physical
scenarios, typically using hybrid systems analysis and novel models of
programming, as advocated in~\cite{DerlerLV12}.

In~\cite{Lee2015},  Lee refers to cyber-physical systems as the ``fundamental intellectual problem of conjoining the engineering traditions of the cyber and the physical worlds'';
he 
also sees models as being central to the study of such cyber-physical
systems, noting  that     
``software abstractions were not created for cyber-physical
systems. They were created to run payrolls''. We agree:  we do not yet have the best abstractions and concepts for
sensor-driven systems. Sensor systems dwell within other systems: water distribution systems, smart grid systems, industrial control systems, environmental systems and all these envelop the sensor systems impacting on them and being impacted by them. One concrete example is of a sensor device that uses vibrations to carry out condition monitoring of an engine; it is responsible for  interpreting the vibration and temperature signals and to quickly inform the actuator to slow the operation of the engine when either the temperature or vibrations reach certain threshold levels. This aims to maximise the lifetime of the engine, but revolution rates cannot fall below the engine efficiency requirements. The sensor device uses a piezoelectric harvester to obtain energy from the engine, yet when the vibrations lower the energy opportunity falls, and a dip in voltage can affect the sensor reading accuracy, cause clock drift etc. The sensor node can provide guarantees that the energy levels are maintained, but is this at the cost of engine longevity? What we have described here consists of many sub-systems, the cyber computing context, the engine itself, its environment, the regulatory framework, etc. The nature and complexity of such hierarchical systems means that no single model can effectively cover everything. Therefore, we agree with other researchers investigating this subject that no one model fits all, however there is a lack of research in articulating how we bring these models together in a comprehensive yet tractable manner; this paper aims to addresses this omission. 

The problem we address is how, 
from a statement of  system requirements and stakeholder concerns, 
and an understanding of system architecture, 
we can derive both models and  questions for  analysis, as well as  interpreting the answers, to assure stakeholders. Work in~\cite{Modelling} reminds us  of the importance of clarity about  {\em model  purpose};  and to  be clear about what a model {\em is} as what it {\em is  not}. To that end
 frames  improve the modelling and analysis process   by  enabling  us to organise models and analysis   and the ways they  contribute
 to assuring system requirements and  stakeholder concerns.   Our goal  is not a formally defined  process,  rather we offer these concepts as an  aid to, or framework for, model development.

\section{Frames of reference}\label{reference}

Stakeholders often have very different (possibly competing) perspectives on the key
abstractions and  assumptions about a sensor-based system. For example developers may be  focussed on the constraints (e.g.~energy consumption) whereas   users may be focussed on the services provided. 
Furthermore, the physical and digital  environments in which  sensor-based systems operate  are inherently complex, and so 
  it can be difficult to  see how the different
  perspectives   can be managed effectively.  
 We define a \emph{frame  of reference}   as a context  in which measurements, judgements and interpretations can be made. 
 Each frame articulates a different perspective and the dimensions and measurements specific to that perspective. 
 Our  definition aligns with standard dictionary definitions such as 
 {\em a set of criteria or stated values in relation to which measurements or judgements can be made; the  overall context in which a problem or situation is placed, viewed, or interpreted; the observer interprets what he sees in terms of his own cultural frame of reference};  we refine the concept further by enumerating the   
following   frames relevant to sensor-driven systems.

\begin{itemize}

\item  \textbf{Geographic:} spatial and topological relationships between sensors. Typically these  may be   
   static networks  (e.g.~because the sensors are fixed to lampposts or in the ground), 
   or      dynamic  networks (e.g.~because they are fixed to  mobile people, animals or
  objects),   or on a fluid surface  or  within a flow  in a pipeline.  There may be many such relationships, one of which may be the data communication between sensors and controller(s),  e.g.~in  star topology, peer to peer,  etc.   

\item  \textbf{Physical:}   aspects  of the   
environment in which sensors operate, be it physical or digital, which may affect system behaviours, e.g.~physical 
aspects such as  temperature, wind speed, may affect  degradation of devices and other system components.

\item\textbf{Failures:} relationships between the components that
  can fail, degrade,  or operate incorrectly, including fail-safe mechanisms and
  redundancies.

\item \textbf{Economic:}  quantitative aspects of resources and their 
  consumption, production and discovery. Example resources include   energy, money,   and 
digital data;   typical measurements include   maintenance costs,    constraints on data buffers  and  communication bandwidth.  

\item  \textbf{Legal:} deontic concepts such as obligations, permissions and responsibilities
  for different system components and human users.

\item \textbf{Security:} vulnerabilities, threats and their
  mitigations, such as access controls preventing unauthorised entry
  to a system and encryption methods that encode data so it can only
  be accessed via keys. These vulnerabilities include those presented
  by multi-tenant architectures and the spatial factors of access, for
  example close-proximity, remote-proximity.

\item  \textbf{Privacy:} anonymity, identity, authentication of
  personally identifiable information, and controls on intended and
  unintended disclosures.
  
\item  \textbf{Social:} communication and interaction  
  between humans involved in the system, and between humans, the
  physical/natural world, and the underlying technologies. Usability and cognitive dissonance 
are key concepts. For example,
  GPS drift can cause cognitive dissonance and lack of trust in a
  system when  a user  believes  they are   stationary yet their GPS coordinates are fluctuating.

\item \textbf{Uncertainty:}  what are the acceptable  bounds of uncertainty for various aspects of the system,  
and how    are the  bounds  qualified, quantified, and related to each other.

\item \textbf{Temporal:} aspects of timing and computation paths that are relevant to system purpose,  from simple temporal orderings to hard real-time constraints, as well  also  expected certainty of the model over time. For
    example weather forecasting becomes less certain the further we
  look into the future, and navigation models become less precise as
  we move away from the position where we last verified a location.

 \end{itemize}
We now turn our attention to the question of sensor-rich systems architectures. 
 
\section{Frames  of function}\label{models}
 
\begin{figure*}[htbp]
  \begin{center}
    \includegraphics[scale=0.85]{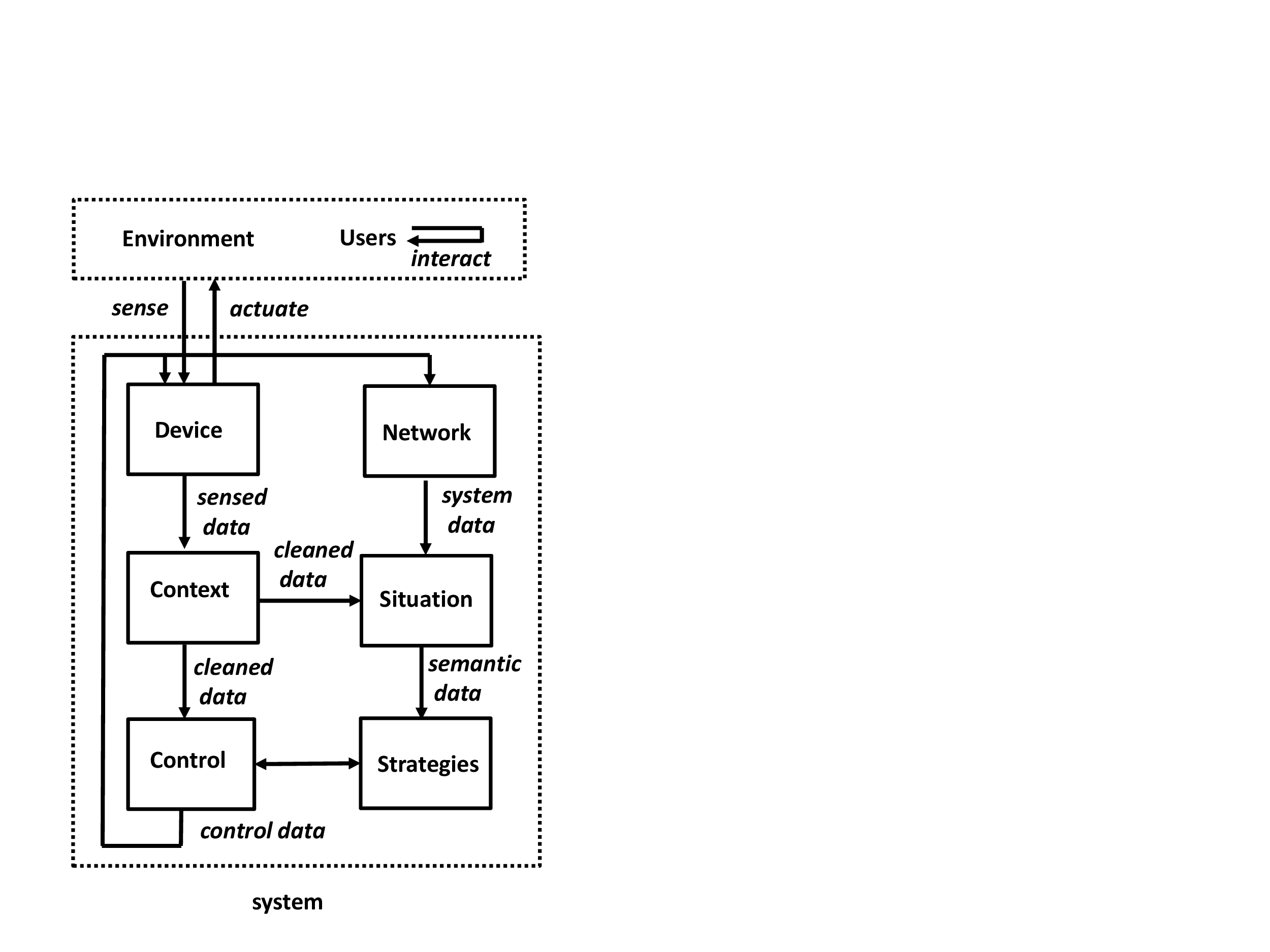}
    \end{center}
  \caption{Typical sensor-driven system   architecture: functional components and data   flows. }
  \label{models_dataflow}
\end{figure*}

  Fig~\ref{models_dataflow} illustrates a typical system software architecture  in a  smart,   sensor-based system: the functional components and  data  flows. Components may relate directly to    physical entities (e.g.~devices) or be conceptual (e.g.~situation, control).  On the left, we have a standard  sense and control loop, whereas on the right we have  network communication, semantic and adaptive/autonomous  capabilities,  and above we have the physical/digital environment and  human users. 
  We define a  \emph{frame  of function} as a component within this architecture, which we describe as follows. (In simple  systems,    some of these   components may  not be present.  )

\begin{itemize}

\item  \textbf{Devices:}     interact directly
with the environment, for  sensing and actuation, they may also have a   housekeeping role. Raw data from   devices are
processed as  context, from which situations  are inferred; 
control and strategies  (and possibly adaptation)   depend on those situations.  
There may be  impacts of the physical environment on devices, e.g.~degradation due to wear and tear, or effect of  temperature on performance.
 
\item \textbf{Context:} refers to  how raw signals from devices
are processed into reusable computer data. This is typically a
syntactic transformation that may, for example, include data smoothing, outlier rejection, 
and the extraction and maintenance of confidence intervals. We often refer to data that 
has been processed in this way as cleaned data. 

\item  \textbf{Control:}   embodies the purpose of the system    through control of  individual devices and the network. For   
  example, control  may include powering down individual devices for periods of time, or changing communication rates.

\item \textbf{Network:}   specifies how and when  devices interact  with each other,
and with the application  (through control).   System data refers to  operational data such as  network status.   
Communications protocols are fundamental  and usually   depend on properties of the underlying sensor topology, e.g.~star topology, peer  to peer,   and properties such as requirements for consistency between adjacent sensors; these properties may be dynamic. 
Sensor topologies may be modelled by metric or topological spaces. Models may also include representation of the impact of the
physical environment on sets of devices and/or communications links.

\item \textbf{Situation:} refers to how data is interpreted
semantically from context, typically by an inference
process~\cite{SituationIdentificationSurvey-10} that is  implemented in
software. For example, a data item might be classified as a known or  a previously undetected event,  or for a given time series, an inference process may infer semantic quantities such as ``getting
hotter'' or ``approaching a dangerous level''.

\item \textbf{Strategies:} are how  the software uses self-reflection (e.g.~adaptation, autonomy) 
to achieve  its
mission and purpose, including node-level, network-level and
application-level adaptations. Strategies may be
tailored to the behavioural envelopes for individual nodes and
ensembles. Behavioural envelopes can change, usually triggered by
situation changes, and adaptation means that devices may be updated,
enhanced or reconfigured. There is usually a close relationship between control and strategies, 
control   typically  depending on   a mixture of    strategies   and  (cleaned) data.

 \item\textbf{Environment:} includes the laws and problem descriptions of aspects of the 
environment, be it physical or digital,    which  affect the function and purpose of the system.
These come from natural
science, engineering, regulations, and so on.  For example, this 
might include  analytical models such as  differential equations, or  statistical distributions, simulations, and
specifications of safe and unsafe thresholds.

\item\textbf{Users:} if human users are involved in the system and interacting with devices (e.g.~wearing sensors) their  behaviour can be considered   as part of the environment. Alternatively,  we may require more explicit  consideration  of   relevant aspects of their  behaviour. 
For example,  this might include interactions between users and devices, or interactions  between users, based, for example, on proxemics --  theories of how people  use interpersonal distance to  understand and mediate   interactions with others \cite{proxemics3}.

\end{itemize}

\section{How and when to use frames}\label{when}

We   propose a  pragmatic,  semi-structured process for employing frames, based    on     plain text. Clarity about    what is in scope, as well  as what is {\em not},  can help  us to focus on the purpose of each model and analysis, and identify any gaps we have overlooked. 
  We have deliberately chosen not to define a formal meta-level framework for modelling: our experience tells us that what is required most urgently is clarity about the purpose and dimensions of a model, rather than   another formal framework that is only used by a small community of experts.      We make  no assumption about the type of model, e.g.~continuous, discrete, stochastic, state-based, event-based, logic-based, agent-based etc., but  do   assume   there are established techniques  and tools for formal analysis.   The following is a guide to using frames.  
\medskip

\noindent Frames are used when 
 \begin{itemize}
\item selecting the modelling formalism and developing a model, 
\item selecting questions  that can be asked of the model and encoding those questions using the analysis techniques for that formalism,   
\item interpreting and communicating  answers from model analysis to stakeholders, 
\item documenting a model for possible future re-use and/or certification,
\item reviewing model coverage:  identifying and explaining gaps in a set  of models,   for example which frames of reference/function are missing and why.   

\end{itemize}
The selection and combination of frames (reference and function) for each model   involves discussion between modellers, analysts, and  systems developers.  While  the primary focus may be a particular set of frames,  a  model may need to have abstract representations or requirements  derived from   one set of    frames,   in order to represent and analyse in more detail another   set. For example, we may require representations  about uncertainty of devices and their   control,  in order to  model uncertainty in the network communications.

\section{Modelling  with frames: examples from a smart water
distribution network}\label{example}


The following  example  is taken from  our experience of working with   smart water
distribution networks 
that monitor  and control pumping, valves, and communication.  In
these  systems,  treated water must be kept moving at sufficient hydraulic
pressure  to maintain the treatment, yet should not put undue pressure on the pipes. Pumping treated water reservoirs to supply
zones and storage tanks consumes most of the energy budget for a
utility.  Smart use of tanks and reservoirs, and shifting pumping schedules
to cheap tariff periods, can result in savings and so impact how water systems remain sustainable. 

 The   purpose   of a
smart water distribution network is to minimise pumping costs, minimise pipe
degradation and leakage, and satisfy customer demands over water
pressure and quality.    The systems includes 
    water reservoir(s), water tanks, flow meters, pressure sensors, motorised valves, and pump(s).
Data communication --  between nodes (sensors) and pump, and from pump to valve --  may be
subject to transmission delays and bandwidth constraints. 

Typical   stakeholder questions 
include: what is lowest pressure that can meet demand and keep water
clean; what is the highest pressure that minimises pipe damage; what
is the minimal data rate that meets legal requirements for reporting
leaks; under the assumption of no failures,  does the proposed pumping schedule   meet customer demand over a 48 hour period;  how resource usage   be reduced , e.g. energy, in delivering water sustainably,  given changing environmental conditions?
 
Modelling   this type of system is challenging because we have    {\em continuous}  behaviour of  fluid flows   and  motorised valves (devices),   {\em stochastic}  behaviour of  digital communication  (network)  and     pipe degradation, and   {\em discrete} behaviour of control software. There are periodic and aperiodic events.  Note there are no  human users nor personal (e.g.~billing) data in the system.    
 
 We describe informally below four   models that are under development   based on {\em Waterbox} -- an   experimental platform~\cite{kartakis2015waterbox} developed at Imperial College. We      list  the primary frames, the modelling and analysis techniques.  Details of these models will be reported in  further technical papers.

  \subsection{Dynamics of fluids and tolerances}\label{model}
  
  \noindent \emph{Frames of reference: Physical, Geographic, Temporal}

 \noindent  \emph{Frames of function: Control, Environment}
 
This model   represents the   dynamics of  fluids and   physical control with  differential equations based on control theory;  embedded in them is the relevant physics of  the physical water system (hydraulics etc.).  In the cyber-physical systems (CPS) community this would be described as representing the physical model.  
The purpose of this continuous model is to investigate under what circumstances  the system control ensures the overall system meets  the  requirements of customer demand  and there are no overflows. Delays in control signals can be as detrimental to the system converging as not being able to detect the state of the water system. The impact of delays (from  communications or other sources) has to  analysed, for each particular controller design.   Specifically, the purpose of each of these models (for each controller design)  is to determine  control {\em tolerances}: what are  the  maximum message  delays and message loss that   can occur and  the system still fulfil requirements of customer demand with no overflows,   for how long   is this  maintained, and what is the degree of disturbance that can be tolerated.  Within these frames we  are not interested in the causes of delays: it is not important whether this is caused by node failure, bandwidth issues or packet loss.
However, the number of devices reporting to a centralised controller can delay its calculations therefore the processing capacity of the controller   is  very relevant.  Considering the  Temporal  frame of reference, there would  not be many changes in the models over time -- except where the water network itself is expanded or trends in demand change due to, for example, a new hospital being built that would cause a impactive change in demand calculations and therefore control schemes.
Analysis of the equations is by simulations,   and typically uses  standard control systems development tools such as Simulink~\cite{simulink}.   

\subsection{Impact of data communications}

\noindent \emph{Frames of reference: Economic, Uncertainty, Failures}

 \noindent  \emph{Frames of function: Network, Devices}
 
Here, the resource in the Economic frame   is data;  the stochastic behaviours  in the Uncertainty frame  arise  from packet loss and delay.   The purpose of this model  is to  investigate different  protocols and  data routing algorithms for data communication in  the network,     the different ways to deal with the  bandwidth and device data buffer constraints, and   when/how individual devices contribute to   delays of   packets. We assume  we have  already determined the tolerances of the control function, namely  maximum message  delays and message loss that the control function can tolerate and still fulfil requirements of customer demand and no overflows etc., in the model above (Section~\ref{model}). 
Most  control systems   utilise a TDMA style MAC protocol (as opposed to CSMA and multi-hop approaches),  because of the tight timing orchestrations that ensure  when  a node sends is more deterministic and therefore packet reception times are more predictable. We examine the use of LoRa, a standard low-powered wide-area   communications to relay data to the base-station;   LoRaWan (its MAC) can provide guarantees of sensor data deliveries due to its scheduled TDMA nature. 
We can describe this as the ``cyber'' side of the CPS equation, and we can model this with  a probabilistic discrete space model.  We   use  probabilistic temporal logic properties to express whether    the tolerances of message delays and loss  will always (or sometimes, and under which circumstances)  be met, and use the model-checker PRISM~\cite{Kwiatkowska:PRISM}, to prove them,  or if not, to indicate the circumstances under which they fail to be met.  

\subsection{Energy consumption}

\noindent \emph{Frames of reference: Economic, Geographic}

 \noindent  \emph{Frames of function: Network and  Strategies}
 
Here, the resource in the Economic frame   is energy and  the aim is to show (or not)  that the battery operational lifetime lasts greater or equal to two years (to minimise the maintenance costs of the sensor network),  and {\em at the same time} the quality of the data is maintained such that the there is at least one data reading available every $15$ minutes. We assume that the network uses low-powered wide-area communications to relay data in a star topology to a single base-station (Geographic frame of reference). Compressive sensing techniques are used to  adapt the sensing rate down when data values indicate that the next sensor reading is not necessary (this technique can interpret data  trends and predict that, for example, temperature will not change more than a degree in the next time slice), which is examined under the Strategies frame of function.  The model is again probabilistic discrete state, but in this case the focus is not on the probabilities \emph{per se}  but on  the       rewards (i.e.~costs)  that are associated with both states and transitions.    Rewards are a feature  of the PRISM~\cite{Kwiatkowska:PRISM} model checker, which is  used to    examine reward based temporal properties that  express assurances about  battery lifetime, while  meeting data sampling requirements.        Note that to offer stronger assurance,  we may in future  add the Situation frame of reference and   make inferences   about energy consumption over sets of observed traces  over time,  and use these to  fine-tune   the rewards.

 \subsection{ Detecting security attacks}
 
 \noindent\emph{Frames of reference: Security}
 
 \noindent \emph{Frames of function: Context, Situation }

Finally,   we consider    how two types of  Security  attack  on the communications network impact the control operation.  Jammed  communication   acts as a delay  and   spoofing can inject false data.   In both cases the result  is  incorrect system (state) and semantic data  that  causes system to function incorrectly. For example,   these attacks    can  cause pipe bursts due to constant actuation causing water pressure hammers,  or can allow water to overflow reservoirs. 
We examine  (cleaned) sensed   data and system in the   Context and  Situation  frames, using machine learning,  to  detect   ``abnormal''  sensing and networks communications.   We can infer stochastic models of clusters of ``normal'' behaviour and variances   from sampled  event  streams, consisting  of  both periodic  data sensing events and aperiodic control events.  We then use the models at runtime, as online monitors  to  detect   ''abnormal''  sensing communication, by comparing  the  observed  data streams from the system with the expected streams.

  \subsection{Model coverage}
  
The simple tables in Figures~\ref{fig:1} and~\ref{fig:2} summarise   the  frames employed    in the  four models above: $1$ refers to model in Section~\ref{model}, etc.  Frames  label the rows:   frames of reference are on the left, frames of function are on the right.  An entry indicates the frame was used in that model, e.g.~Geographic frame of reference was employed in models 1 and 3, the Environment frame of function was employed in model 1.  
 The  Social and  Privacy frames of reference are omitted as they are  not relevant to this system. We do not expect a uniform  distribution of frames and it is not surprising that  the Geographic and Economic frames of reference  and  Network frame of function are addressed in more than one model --  these are key frames for this system.    We can see quickly that we have not addressed the Legal frame of reference. This is deliberate, for this system, because  burst detection is straightforward  (e.g. from flows) and sub-second notification is not required.



\begin{figure}[h]
\begin{center}
  \begin{subfigure}[b]{0.4\textwidth}
    
   \begin{tabular} {   r | c |   } 
 Frame of reference & model   \\ \hline
Geographic  & 1, 3     \\   
Physical   &1   \\ 
Economic  & 2, 3   \\ 
Security   &4   \\ 
Uncertainty   &2   \\ 
Temporal  &1  \\
Failures & 2   \\
Legal  & --     \\  
\end{tabular} 
    
    \caption{Frames of reference.}
    \label{fig:1}
  \end{subfigure}
  \begin{subfigure}[b]{0.4\textwidth}
   \begin{tabular} {   r | c |   } 
 Frame of function & model   \\ \hline
Environment  & 1   \\   
Device   &2   \\ 
Context  & 4   \\ 
Control   &1   \\ 
Network   &2, 3   \\ 
Situation  &4  \\
Strategies & 3   \\
\end{tabular}
    
    \caption{Frames of function.}
    \label{fig:2}
  \end{subfigure}
  \caption{Frames of reference (left) and function (right) employed   in models 1--4.}
  \end{center}
\end{figure}

 \section{Discussion}\label{discussion}

The  variety of   models required for this one  system  demonstrates the breadth and complexity  of     sensor-driven networked systems and  their requirements.   In this case no   single  model would be tractable,  or comprehensible:    each model has a purpose and analysis of that  model    contributes to our overall  understanding and trust that the  system fulfils {\em its} requirements. Frames help   us to distinguish and articulate the purpose  of each model, and also to articulate the absence of a model.  

We recognise that models are just one, albeit important, component of assurance  of sensor-driven systems: modelling and analysis needs to be coupled   with software development, testing and analysis of actual deployments. While techniques for context and situation recognition and inference have, and are, being
researched extensively, especially in the context of ``big data'', we suggest there are several new challenges specific to sensor-based system  modelling. For example:

\begin{itemize}
 \item How can historical data be used to calibrate models that have not been inferred from data,   for the analysis of system play-back, and also to aid predictions of future system behaviour? 

\item  How can we use model predictions as inputs to control,  and use sensed contexts and situations to calibrate model 
  predictions?  For example, can we use model predictions  
 to modify system goals,  to drive sensor change, and detect data anomalies? 

\item 
 Over time,  context  and situations that have been observed and  inferred may be compared with those
 predicted by standard   models based on laws of physics,  e.g.~observed (sensed)  water levels are compared with  predicted water levels.   But we may find that the observed levels are not as predicted.   
How should we modify the  (possibly long established)  
 physics  models when they do not align with actual sensed data?  
 
\item What can we learn from sensor-driven systems that fail, as well
as succeed? Too often, only successful applications are recorded.    What is the role of modelling in failed systems?  
  A good example of a failed application is the remarkably
honest assessment of a 150 wireless sensor node application for
precision agriculture reported in~\cite{Langendoen06}.  A culture of
reporting failures, and analyses thereof (e.g.~the recent FAILSAFE workshop \cite{failsafe})  will only strengthen the
role  for improved modelling and also for sensor-driven systems.
\end{itemize}

\section{Related work}\label{related}
   
The problem of identifying and dealing with multiple perspectives, or views, of a system, particularly during software development, is well known in software engineering.   We attempt here to situate our work in that context, noting that we do not give a comprehensive literature review, but rather select key related work.  

Nearly  thirty years ago Somerville  introduced {\em viewpoint analysis}~\cite{Sommerville}, where a viewpoint is defined informally as   a ``way of looking at the system''.  Viewpoints are very wide ranging:   there may be tens or hundreds of viewpoints for a system, the exact number and nature depends on the system.    There is some categorisation, e.g.~there are functional viewpoints and non-functional viewpoints. Very loosely, we can say that our frames of  function correspond   to the former and our frames of  reference to the latter. Further comparison is difficult because Sommerville’s viewpoints  cover  all the breadth  (and depth) of software systems;  we are concerned with a specific domain (sensor-driven systems),  and so we can be much more prescriptive about our frames.    

Subsequent research on views has been more rigorous, such as contract-based design and multiple viewpoints~\cite{view1,view2,view3},  which    aim to support distributed design of different aspects of the system; the key results are generic mathematical models that formalise  contract theories and meta-theories.  We have deliberately chosen not to formalise processes, e.g.~how to integrate models,  nor to introduce   new modelling techniques, rather our aim has been a pragmatic,  semi-structured process to improve model development and interpretation and communication of results.  

Several researchers have recognised the difficult problem of {\em framing}, for example problem frames~\cite{Jackson} and experimental frames~\cite{Zeigler,Daum}.   The former presents   a set of concepts  to be used when gathering requirements and creating specifications for   software, these include specific frames of required behaviour, commanded behaviour, information display, etc.  Again, while these are   useful in the context of general software development, they are focussed on requirements and design,  and don’t meet our specific needs of  articulating model purpose  for    sensor-driven systems that operate  in a  cyber-physical world.  The latter is defined specifically for  discrete event simulation  models  and addresses the problem of   characterising a set(s)  of circumstances under which  a model is to be observed or experimented with.    This includes specifying the mean rate of arrivals,  seeds for the random number generators etc.  This provides  a useful, but narrow lens with which to view  models, and apart from assuming the model is event based, it disregards any domain specific aspects such as what the events represent.  We   expect   experimental circumstances to be  part of the dimensions of a frame, or pair of (reference  and  function) frames.   For example, rates of  communication bandwidth would be specified within the economic  and network frames, rates of sensor  failures within the failure and  device frames, and      message loss rates within the     failures and network  frames.  Finally, there are few explicit concepts for  modelling sensor-driven systems, apart from those we mentioned earlier for pervasive computing and cyber-security, though we note that   ~\cite{Tsigkanos}  classifies  properties as  local spatial (entities are in pre-defined structural  patterns), global spatial (entities are arbitrarily distributed in space), and temporal, which loosely correspond to our geographic and temporal frames of reference.

Can our concepts add  new insights into  existing models of sensor-based systems? 
For example, the Ptides and Ptolemy models described  in~\cite{DerlerLV12} and~\cite{Lee2015}   refer  to  interplays between 
 physical models (i.e.~Environment), sensors, actuators (i.e.~Devices),  computation (i.e.~Control), and Network communication.   The  Failures and Temporal frames of reference  are predominant, with an emphasis on timing and scheduling.  Additionally, in \cite{DerlerLV12}
a   modal,   hybrid model  in Ptolemy II is  proposed   for improved  adaptation, we would classify this  as  focussing on the  Strategies frame.  
 Future work  could  determine how our frames    help with  modifications to these systems and/or how the models could  be extended to incorporate assurance over other frames of reference.

\section{Conclusions}\label{conc}

Sensor-driven systems will become increasingly significant over the
coming decades, supporting evidence-based decision-making in the face
of global challenges such as environmental change, food production, internet of things and
autonomous vehicles.  But progress will be undermined
by our inability to assure both decision-makers and users of the
integrity and timeliness of the information being provided and the
decisions taken. Without such assurance, increasingly ``smart''
infrastructures will become unpredictable and unacceptable, and the
ability of state, scientific, and industrial actors to leverage the
benefits of sensing and autonomy will be severely restricted.


Modelling is a crucial part of establishing trust, and models are    applicable not only at design time, but also during deployment, as a system is running.  
We have introduced two new concepts: frames of reference   and
frames  of function, which provide an abstract methodology for design and
and  development. They  improve the modelling and analysis process by  helping   us to distinguish and articulate the purpose  of each model, which in turn   helps us  to deal with, communicate, and assure  stakeholders about    the competing demands of understanding the  sensed world, 
the computation, communication, and self-reflection within a sensor based system,  and the ways in which the devices interact with sensed world.

We have outlined how these concepts are being
employed in an  example  system    we are working on;
demonstration of the full benefits of these models is the focus of further technical papers.
Future work will include  further demonstration and development of frames of reference and function, including  the concrete languages and ontologies needed,  through case studies.  The case studies will be based on increasingly complex end-user applications that cover a wide range of frames  and   include sensor-driven systems that fail, as well as succeed.



\subsection*{Acknowledgments}
This work is supported by the {\em Science of Sensor System Software} programme grant,  funded by the EPSRC (Engineering and Physical Sciences Research Council),  EP/N007565/1.

\bibliographystyle{abbrev}

\begin{thebibliography}{}

\bibitem[fai, ]{failsafe}
First {A}{C}{M} international workshop on the engineering of reliable, robust,
  and secure embedded wireless sensing systems ({F}{A}{I}{L}{S}{A}{F}{E}).
\newblock \url{http://wp.doc.ic.ac.uk/failsafe}.

\bibitem[Benford et~al., 2016]{Savannah}
Benford, S., Calder, M., Rodden, T., and Sevegnani, M. (2016).
\newblock On lions, impala, and bigraphs: modelling interaction in
  physical/virtual spaces.
\newblock {\em ACM Transactions on Human-Computer Interaction (TOCHI)},
  23:9:1--9:56.

\bibitem[Benveniste et~al., 2007]{view3}
Benveniste, A., Caillaud, B., and Passerone, R. (2007).
\newblock {A Generic Model of Contracts for Embedded Systems}.
\newblock {\em INRIA Rapport de Recherche No. 6214}.

\bibitem[Calder et~al., 2015]{MAM2015}
Calder, M., Gray, P., and Unsworth, C. (2015).
\newblock Is my configuration any good: checking usability in a sensor-based
  activity monitor.
\newblock {\em Innovations in Systems and Software Engineering}, 11:131--142.

\bibitem[Calder~et al, 2018]{Modelling}
Calder~et al, M. (2018).
\newblock Computational modelling for decision-making: where, why, what, who,
  and how.
\newblock {\em Royal Society Open Science}, 5:172096.

\bibitem[Coutaz et~al., 2005]{ContextIsCentral}
Coutaz, J., Crowley, J., Dobson, S., and Garlan, D. (2005).
\newblock Context is key.
\newblock {\em Comm. ACM}, 48(3):49--53.

\bibitem[Daum and Sargent, 2001]{Daum}
Daum, T. and Sargent, R.~G. (2001).
\newblock {Experimental Frames in a Modern Modeling and Simulation System}.
\newblock {\em IIE Transactions}, 33(3):181--192.

\bibitem[Derler et~al., 2012]{DerlerLV12}
Derler, P., Lee, E.~A., and Sangiovanni{-}Vincentelli, A.~L. (2012).
\newblock {Modeling Cyber-Physical Systems}.
\newblock {\em Proceedings of the {IEEE}}, 100(1):13--28.

\bibitem[Fang and Dobson, 2013]{ucami-13}
Fang, L. and Dobson, S. (2013).
\newblock {In-network Sensor Data Modelling Methods for Fault Detection}.
\newblock In {O \lq Grady}, M. and et~al, editors, {\em Evolving Ambient
  Intelligence}, pages 176--189. Springer International Publishing.

\bibitem[Fisher et~al., 2013]{FisherDW13}
Fisher, M., Dennis, L.~A., and Webster, M.~P. (2013).
\newblock {Verifying Autonomous Systems}.
\newblock {\em Comm. {ACM}}, 56(9):84--93.

\bibitem[Hennicker and Ludwig, 2012]{view2}
Hennicker, R. and Ludwig, M. (2012).
\newblock { View-Based Development of a Simulation Framework for
  Multi-disciplinary Environmental Modelling }.
\newblock volume 7539 of {\em LNCS}.

\bibitem[Jackson et~al., 2017]{Jackson:2017}
Jackson, G., Kartakis, S., and McCann, J.~A. (2017).
\newblock Accurate models of energy harvesting for smart environments.
\newblock {\em 2017 {IEEE} International Conference on Smart Computing,
  {SMARTCOMP} 2017, Hong Kong, China, May 29-31, 2017}, pages 1--6.

\bibitem[Jackson, 2001]{Jackson}
Jackson, M. (2001).
\newblock {\em Problem Frames: Analysing and Structuring Software Development
  Problems}.
\newblock Addison-Wesley.

\bibitem[Kamal et~al., 2014]{DobsonACM2014}
Kamal, A. R.~M., Bleakley, C.~J., and Dobson, S. (2014).
\newblock {Failure Detection in Wireless Sensor Networks: A Sequence-based
  Dynamic Approach}.
\newblock {\em ACM Transactions on Sensor Networks}, 10(2):35:1--35:29.

\bibitem[Kartakis et~al., 2015]{kartakis2015waterbox}
Kartakis, S., Abraham, E., and McCann, J.~A. (2015).
\newblock Waterbox: A testbed for monitoring and controlling smart water
  networks.
\newblock In {\em Proceedings of the 1st ACM International Workshop on
  Cyber-Physical Systems for Smart Water Networks}, pages 8:1--8:6. ACM.

\bibitem[Konur et~al., 2014]{ScatterboxVerification}
Konur, S., Fisher, M., Dobson, S., and Knox, S. (2014).
\newblock Formal verification of a pervasive messaging system.
\newblock {\em Formal Aspects of Computing}, 26(4):677--694.

\bibitem[Kwiatkowska et~al., 2011]{Kwiatkowska:PRISM}
Kwiatkowska, M., Norman, G., and Parker, D. (2011).
\newblock {PRISM} 4.0: Verification of probabilistic real-time systems.
\newblock In {\em Proc. 23rd Intl Conf. on Computer Aided Verification
  (CAV'11)}, volume 6806 of {\em LNCS}, pages 585--591. Springer.

\bibitem[Langendoen et~al., 2006]{Langendoen06}
Langendoen, K., Baggio, A., and Visser, O. (2006).
\newblock {Murphy Loves Potatoes: Experiences from a Pilot Sensor Network
  Deployment in Precision Agriculture}.
\newblock In {\em Proc. 20th International Parallel and Distributed Processing
  Symposium (IPDPS)}.

\bibitem[Lee, 2015]{Lee2015}
Lee, E.~A. (2015).
\newblock {The Past, Present and Future of Cyber-Physical Systems: A Focus on
  Models}.
\newblock {\em Sensors}, 15(3):4837--4869.

\bibitem[S. et~al., 2011]{proxemics3}
S., G., Marquardt, N., Ballendat, T., Diaz-Marino, R., and Wang, M. (2011).
\newblock Proxemic interactions: The new ubicomp?
\newblock 18(1):42--50.

\bibitem[Simulink, ]{simulink}
Simulink.
\newblock \url{https://uk.mathworks.com/help/simulink}.

\bibitem[Sommerville, 1992]{Sommerville}
Sommerville, I. (1992).
\newblock {\em Software Engineering, fourth edition}.
\newblock Addison Wesley.

\bibitem[Tsigkanos et~al., 2017]{Tsigkanos}
Tsigkanos, C., Kehrer, T., and Ghezzi, C. (2017).
\newblock Modeling and verification of evolving cyber-physical spaces.
\newblock {\em Proc. 11th Joint Meeting on Foundations of Software Engineering,
  {ESEC/FSE}, 2017}, pages 38--48.

\bibitem[von Hanxleden et~al., 2012]{view1}
von Hanxleden, R., Lee, E.~A., Motika, C., and Fuhrmann, H. (2012).
\newblock Multi-view modeling and pragmatics in 2020: Position paper on
  designing complex cyber-physical systems.
\newblock volume 7539 of {\em LNCS}.

\bibitem[Ye et~al., 2012]{SituationIdentificationSurvey-10}
Ye, J., Dobson, S., and McKeever, S. (2012).
\newblock Situation identification techniques in pervasive computing: a review.
\newblock {\em Pervasive and Mobile Computing}, 8(1):36--66.

\bibitem[Ye et~al., 2016]{Abnormal16}
Ye, J., Stevenson, G., and Dobson, S. (2016).
\newblock Detecting abnormal events on binary sensors in smart home
  environments.
\newblock {\em Pervasive and Mobile Computing}, 33:32--49.

\bibitem[Zeigler, 1976]{Zeigler}
Zeigler, B. (1976).
\newblock {\em Theory of Modelling and Simulation}.
\newblock Wiley.

\end{thebibliography}
\newcommand{\BIBentryALTinterwordspacing}{}
\newcommand{\BIBentrySTDinterwordspacing}{}

\end{document}